\documentclass[final
    ,final            % use final for the camera ready runs
  ]
  {aipproc}

\layoutstyle{6x9}
\usepackage{graphicx}

\begin{document}

\title{The phases of strongly interacting matter in heavy ion 
collisions}


\author{P\'eter L\'evai}{
  address={RMKI Research Institute for Particle and Nuclear Physics \\
           P.O. Box 49, Budapest 1525 Hungary}
}

\begin{abstract}
In ultrarelativistic heavy ion collisions the produced 
high temperature, high energy density
state will cross different phases of the strongly interacting matter. 
The original idea of quark-gluon plasma formation has been evolved and
the weakly interacting gaseous state of massless partons
has been replaced by the picture of strongly interacting
massive constituent quarks.
Experimental results at SPS and RHIC supports this idea.
We discuss the phases at LHC energies.
\end{abstract}

\maketitle

%%%%%%%%%%%%%%%%%%%%%%%%%%%%%%%%%%%%%%%%%%%%
%% MAINMATTER
%%%%%%%%%%%%%%%%%%%%%%%%%%%%%%%%%%%%%%%%%%%%

\section{Introduction}

Ultrarelativistic heavy ion experiments at CERN Super Proton Synchrotron (SPS)
and BNL Relativistic Heavy Ion Collider (RHIC)
focus on the production of the Quark Gluon Plasma (QGP) state, 
which is described by the theory of Quantum Chromodynamics (QCD).
The formation of this state was suggested 25 years ago~\cite{QGP0}: 
with increasing 
collisional energy the produced energy density in the center of heavy ion
collisions may reach a critical value, where proton and neutron
degrees of freedom are melted and quarks and gluons determine
the properties of the produced hot matter~\cite{Muller85,qgp}. 
Theorists have been expected a first order {\it ``nuclear matter} 
$\leftrightarrow$ {\it QGP''} phase transition with large latent heat
at finite density and second order phase transition at zero baryon density.
This latent heat is connected to the disappearance of the 
massive hadronic states and the formation of a weakly interacting plasma state
of massless partons. With increasing 
collision energy simply the temperature of this plasma state 
would have been increasing.

However, in the mid-90's Pb+Pb collisions at CERN SPS 
did not show clearly this phase transition and the formation
of the weakly interacting QGP, although the necessary energy 
concentration has been reached.
'Compelling evidences' of the transition
(strangeness enhancement, melting of charmed mesons, etc.)
have been collected~\cite{CERN2000},
but the properties of the produced quark matter were not identified.
Experimental data forced to replace the bulk
description of the above phase transition with microscopical models.
Quark coalescence became one of the most successful idea~\cite{ALCOR}. 
This model is based on quasi-particles, namely
massive constituent quarks~\cite{Goren95,LH98}, 
and the first order phase transition was successfully replaced 
by a cross over transition with negligible latent heat. This 10 years old
finding is supported by recent lattice-QCD results at finite 
baryon densities~\cite{FodKa02}, where the 
first order phase transition ends at finite temperature and density.
The experimentally accessible regions belong to the 
cross over phase transition region. The presence of
quasi-particles enriched the phase structure at high $T$.
\newpage

\section{Strongly interacting matter and quasi-particles}

The analysis of SU(3) lattice QCD results on equation of state in a
thermodynamically consistent quasi-particle picture~\cite{LH98}
yielded a description of a deconfined state, where
the basic degrees of freedom are quarks, antiquarks and gluons
with the usual color and spin
degrees of freedom (e.g. transverse gluons), but
with an effective thermal mass and width,
 generated by many-body interactions.
The consequences are the following:
\begin{enumerate}
\item
The effective thermal masses are large around the critical temperature $T_c$,
and they are linearly increasing with the temperature
(e.g. in case of $N_f=2$ one obtains
$M_{g,\infty}(T_c) \approx 500 \ MeV$ and
$M_{q,\infty}(T_c) \approx 330 \  MeV$, see Ref.~\cite{LH98}).
\item
The effective strong coupling constant, $\alpha_{eff}$, is large around $T_c$,
the system is strongly coupled. In parallel, the effective width of the
quasi-particles is very small, $\Gamma \sim g^2T \log ({1/g})$, thus 
massive quarks are well-defined objects around $T_c$.
\item
The gluon number  is suppressed, because
massive gluons are  heavier than the massive quarks.
This suppression is large around the critical temperature
(for $N_f=2$ the ratio is
$N_g/(N_q+N_{\overline q})\approx 1/3$ around $T_c$), thus quarks and 
antiquarks dominates the matter. We will call this state as
Quark-Antiquark Plasma (QAP).
\item
The QAP phase is relatively dilute:
the densities of the massive quarks and gluons are much
smaller than in the massless case (one obtains factor 2.5 for quarks and
factor 5 for the gluons in case of $N_f=2$).
\item
The speed of sound is large around the critical temperature, 
$c_s^2(T_c) \approx 0.15$,
which implies fast expansion and fast dynamical evolution of the QAP phase.
\end{enumerate}

Such a massive quasi-particle picture was successfully applied
to describe the soft particle production at CERN SPS and 
BNL RHIC~\cite{ALCOR,ALCOR2,MICOR}. Even more, high-$p_T$ data were reproduced
successfully by quark coalescence at RHIC energies, 
explaining the anomalous antiproton/pion ratio~\cite{HGF1,HGF2,HGF3} 
and the scaling of elliptic flow~\cite{MoV}.

Strongly interacting quarks and antiquarks can form color mesonic 
bound states inside the deconfined matter~\cite{ShuryZah04},
and further complicated objects may appear, thanks to the strong
effective coupling constant between massive quarks and gluons. 
The presence of such a color bounded complex objects can be 
responsible for the appearance
of a liquid-like behavior of the QAP phase around the critical
temperature. Furthermore, these heavy compounds are able to drag 
the even heavier
charm quarks and generate a common quark flow, as it was observed at 
RHIC~\cite{MICOR,Akiba05}.

As the temperature is increasing, the effective strong coupling
constant is decreasing, finally it overlaps with its perturbative QCD
value around $T\sim 3 \, T_c$. In parallel, the width of the quasi-particles
is increasing: when $\Gamma \sim M$, then quasi-particles become
ill-defined, they split into the original degrees of freedom,
namely massless quarks and gluons. 
Since this is a continuous change in the properties
and the number of degrees of freedom does not change, there is a smooth
phase transition between QAP and QGP, the strongly interacting 
QAP phase is the low temperature manifestation of the weakly
interacting QGP.
This smooth change is expected around
$T=400-450$ MeV. At RHIC we may reach higher temperature in the early stage
of the heavy ion collisions, however
the cooling and expanding QGP state will be converted into QAP and
hadronize from this state. This is the reason, why coalescence
models work successfully at RHIC energies.

\newpage

\begin{figure}[ht]
\resizebox{155mm}{185mm}{\includegraphics[height=180mm]{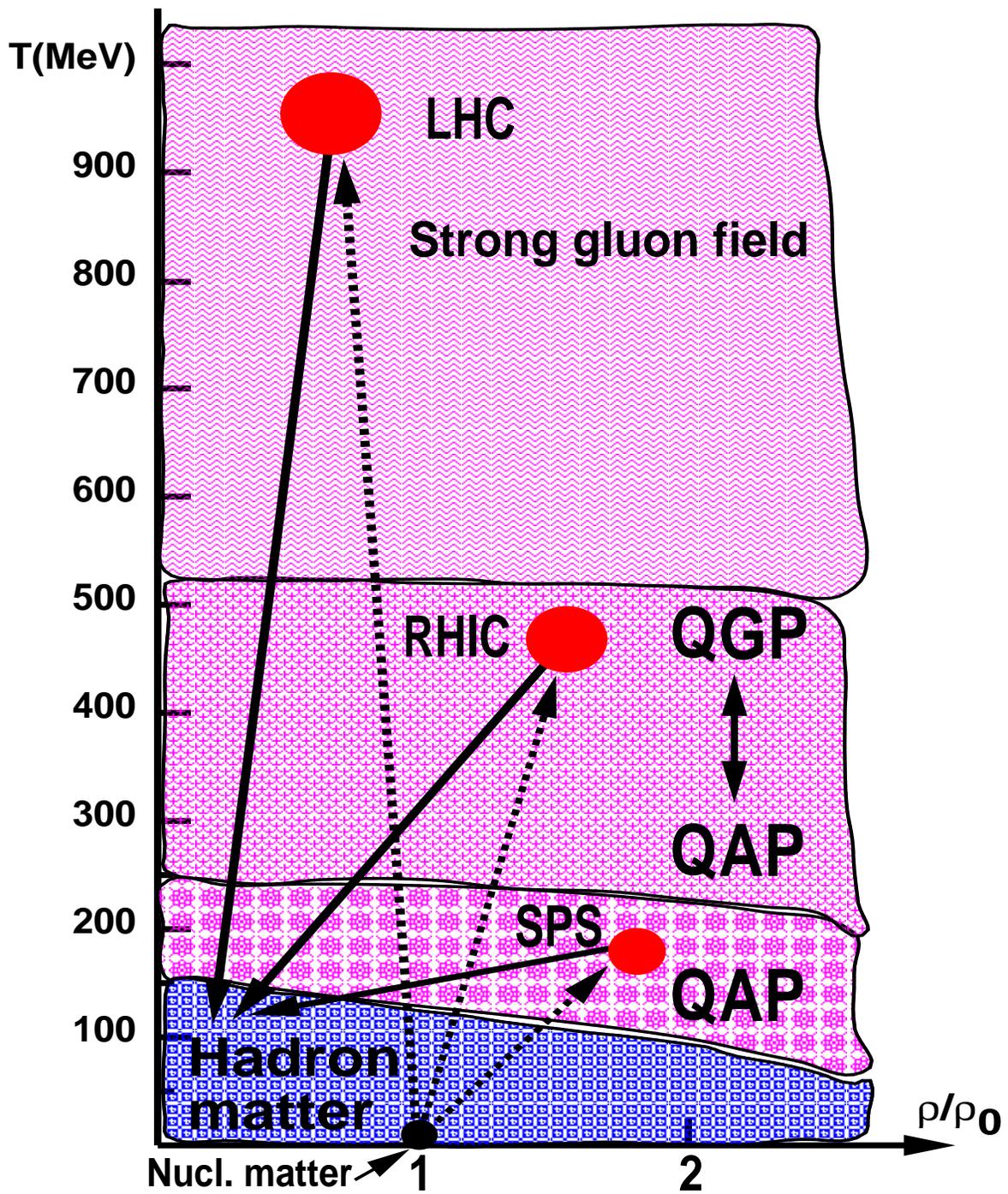}}
 \caption{The different phases of the strongly interacting matter.
     The dotted and solid lines indicate the 
     compression and expansion path of the heavy ion
     collisions at different energies. }
\end{figure}

\newpage

At LHC energies we can reach such a high energy concentration, when
the gluons will dominate the produced matter and a Yang-Mills black body
radiation may appear in heavy ion collisions. This phase is very
interesting, because at high gluon density 
we expect the formation of a classical gluon field and
special saturation behaviors can be studied experimentally.
It is important to investigate, if we can reach such a high
temperature in real heavy ion collisions, 
because direct particle production from strong
(abelian and non-abelian) fields may generate much lower 
temperature~\cite{Skokov05}.

Figure 1 summarizes the discussed phases in a 
temperature---density plot. In fact there are no sharp boundaries
between the different phases, they smoothly overlap with each others.
The dotted and solid 
lines indicate the path in heavy ion collisions at different energies.
To verify our picture, we can consider the results of 
jet-tomography analysis~\cite{GLV}
about the measured color charge densities at different energies.
 
Figure 1 displays static phases of infinite matter. Recently 
we are investigating those properties of the strongly interacting
matter, which can be verified and determined from
microscopical models, e.g. heating, viscosity
in molecular dynamics simulations~\cite{Hartmann05}.

\begin{theacknowledgments}
The author thanks for the financial support of the OTKA
         Grants T043455 and T049466.
\end{theacknowledgments}

\end{document}